\documentclass[12pt,dvips]{article}
\usepackage[dvips]{graphicx}
\usepackage{graphicx}
\usepackage{epsfig}
\usepackage{cite}

\graphicspath{{plots/}}
\DeclareGraphicsExtensions{.eps.gz,.eps,.ps,.ps.gz}

\parskip=\itemsep               
\newcommand{\lwig}{\mbox{\,\raisebox{.3ex}
    {$<$}$\!\!\!\!\!$\raisebox{-.9ex}{$\sim$}\,}}
\newcommand{\gwig}{\mbox{\,\raisebox{.3ex}
    {$>$}$\!\!\!\!\!$\raisebox{-.9ex}{$\sim$}}\,}
\newcommand{\lambdabar}{{\hbox{$\lambda_e$\kern-1.9ex\raise+0.45ex\hbox{--}
\kern+0.2ex}}}
\newcommand{\bbar}{\overline}

\newcommand{\lf}{\left}
\newcommand{\rg}{\right}

\newif\ifhepph

\hepphtrue

\ifhepph\date{\empty}\fi

\title{
\ifhepph{\normalsize
\rightline{DESY 06-069}}\fi
\vskip 1cm 
\bf\boldmath
Instantons in Lepton Pair Production  
       \vspace{21mm}} 
\author{
Arnd Brandenburg\footnote{Present address: Genedata AG, Maulbeerstra\ss e 46, 
CH-4016 Basel, Switzerland}, 
Andreas Ringwald\\[4mm] 
Deutsches Elektronen-Synchrotron DESY, Hamburg, Germany\\[8mm]
Andre Utermann\\[4mm]
Department of Physics and Astronomy, Vrije Universiteit Amsterdam,\\ The Netherlands}

\begin{document}
\begin{titlepage} 
  \maketitle
\begin{abstract}
We consider QCD instanton-induced contributions to 
lepton pair production in hadron-hadron collisions. We  
relate these contributions to those known from deep inelastic
scattering and demonstrate that they 
can be calculated reliably for sufficiently large momentum transfer. 
We observe that the instanton contribution to the angular distribution of the 
lepton pairs at finite momentum transfer strongly violates the 
Lam-Tung relation -- a relation between coefficient functions of the angular distribution 
which is valid within the framework of ordinary perturbation theory.
The drastic violation of this relation, as seen in experimental data, 
might be related to such instanton-induced effects.  
\end{abstract}


\thispagestyle{empty}
\end{titlepage}
\newpage \setcounter{page}{2}

\section{\label{intro} Introduction}

The Standard Model of electroweak (Quantum Flavor Dynamics (QFD)) and 
strong (QCD) interactions is extraordinarily successful. This success is largely based
on the possibility to apply ordinary perturbation theory to the 
calculation of hard, short-distance dominated scattering processes, since
the relevant gauge couplings are small. 
Certain processes, however, can not be described by ordinary
perturbation theory, no matter how small the gauge coupling is. These processes 
are associated with axial anomalies~\cite{Adler:gk} and 
manifest themselves as anomalous violation of baryon plus lepton  number ($B+L$) in 
QFD and chirality ($Q_5$) in QCD~\cite{'tHooft:1976up}. 
They are induced by topological fluctuations of the non-Abelian gauge
fields, notably by instantons~\cite{Belavin:fg}.

A number of non-perturbative issues in the Standard Model can be understood 
in terms of such topological fluctuations and the associated anomalous processes. 
On the one hand, QCD instantons seem to play 
an important role in various
long-distance aspects of QCD, such as providing a possible solution to 
the axial $U(1)$ problem~\cite{'tHooft:1976up} or being at work in
chiral symmetry breaking~\cite{Shuryak:1981ff}. 
In QFD, on the other hand, 
analogous topological fluctuations of the gauge fields and
the associated $B+L$ violating processes are very important at high 
temperatures~\cite{Klinkhamer:1984di} and have therefore a crucial impact on the 
evolution of the baryon and lepton asymmetries of the universe~\cite{Rubakov:1996vz}.

Are manifestations of such topological fluctuations also directly
observable in high-energy scattering processes?  This question has
been seriously considered in the late 1980's, originally in the
context of QFD~\cite{Aoyama:1986ej}. But, despite considerable
theoretical~\cite{McLerran:1989ab} and
phenomenological~\cite{Farrar:1990vb} efforts, the actual size of the
cross-sections in the relevant, tens of TeV energy regime was never
established (for recent attempts, see Ref.~\cite{Ringwald:2002sw}).
Meanwhile, the focus switched to quite analogous QCD instanton-induced
hard scattering processes in deep inelastic
scattering~\cite{Balitsky:1993jd}, which are calculable from first
principles within instanton-perturbation theory~\cite{Moch:1996bs},
yield sizeable rates for observable final state signatures in the
fiducial regime of the latter~\cite{Ringwald:1998ek,Ringwald:1999jb},
and are actively searched for at HERA~\cite{Adloff:2002ph}.
Moreover, it has been argued that larger-size QCD instantons, beyond
the semiclassical, instanton-perturbative regime, may well be
responsible for the bulk of inelastic hadronic processes and build up
soft diffractive scattering~\cite{Kharzeev:2000ef}.  It was emphasized
for the first time in Ref.~\cite{Schrempp:2005vc} that single photon or single
$W$ production at large transverse momentum offers a possibility to
study QCD instanton-induced effects from first principles at the LHC.
Unlike the processes considered in the present paper,
the dominant subprocess for this dedicated instanton search at high
energies at the LHC~\cite{Schrempp:2005vc,sp} is induced by gluon fusion, e.g.
$gg\to V+X,\,V=\gamma^\ast\to \ell^+\ell^-$. 
 Moreover, the kinematical region is remarkable
different from our region of interest, i.e. the available transverse
momenta and virtualities are significantly larger than those we
concentrate on throughout this paper.

\begin{figure}
\begin{center}
  \epsfig{file=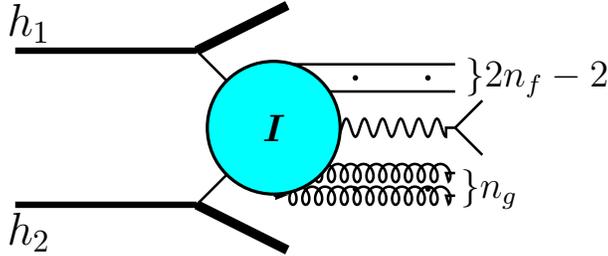,width=8cm}
\caption[dum]{\label{drellyan}
  QCD instanton-induced contribution to lepton pair production in hadron-hadron
  collisions, 
  ${h}_1+{h}_2 \rightarrow (n_f-1) \left[\bar q_R + q_R\right] + \ell^+ + \ell^- 
   + n_g\,g + X$, corresponding to $n_f$ 
  light flavours.} 
\end{center}
\end{figure}

In this paper, we consider QCD instanton-induced contributions to
lepton pair production in hadron-hadron collisions\footnote{This is
  often called the Drell-Yan process~\cite{Drell:1970wh}.  Instanton
  contributions to this process have been first discussed in
  Ref.~\cite{Ellis:1978pe} at a qualitative level.} (cf.
Fig.~\ref{drellyan}). We relate these contributions to the ones
previously calculated for deep inelastic
scattering~\cite{Moch:1996bs}, thereby demonstrating that the former
-- like the latter -- can be calculated from first principles. In
particular, as already emphasized in Ref.~\cite{Schrempp:2005vc}, the
typical inverse hard transverse momentum scale $q_\perp^{-1}$ in
lepton pair production provides a dynamical infrared cutoff for the
instanton size parameter $\rho$, thereby allowing for a controlled
semiclassical approximation, which rests on the smallness of the QCD
coupling at the effective momentum scale $1/\langle\rho\rangle$:
$\alpha_s(1/\langle\rho\rangle )\ll 1$.  Hence, in addition to deep
inelastic scattering, lepton pair production at large transverse
momentum may be viewed as a distinguished process for studying
manifestations of QCD instantons.

We put special emphasis on the angular distribution of the lepton pairs at finite 
momentum transfer. We observe that the instanton contribution strongly violates the 
Lam-Tung relation~\cite{Lam:1980uc} between coefficient functions of the angular distribution, 
which has been verified within the framework of ordinary perturbation theory -- 
the QCD improved parton model -- up to $\mathcal O (\alpha_s^2)$~\cite{Brandenburg:1993cj,Mirkes:1994dp} 
and even holds for the inclusion of parton transverse momentum and soft gluon
effects~\cite{Chiappetta:1986yg,Boer:2006eq}.  
Indeed, it has been argued that the drastic violation of 
this relation, as seen in experimental data~\cite{Falciano:1986wk,Guanziroli:1987rp,Conway:1989fs}, 
might be due to a non-trivial structure of the QCD vacuum~\cite{Brandenburg:1993cj},  
and in particular could be related to instanton-induced
effects~\cite{Boer:2004mv}. 

The outline of this paper is as follows: In Sect.~\ref{dis2drellyan}
we introduce the instanton-induced contribution to lepton pair
production. Afterwards we review the known results in the related
process in DIS in an instanton background. The crucial part of this
section is the continuation of these results to hadron collisions which
leads us to the photon production tensor on partonic level. In
Sect.~\ref{angular} we will use these results to calculate the angular
distribution of the produced leptons on the partonic
(Sect.~\ref{partonlevel}) and the hadronic (Sect.~\ref{hadron}) level.
In Sect.~\ref{gluons} we give an outlook on the inclusion of multi
gluon processes which lead to an enhancement of the
instanton contributions.  We present our conclusions in
Sect.~\ref{conclusions}.

\section{\label{dis2drellyan} From deep inelastic scattering to lepton pair production}

\begin{figure}[t]
\begin{center}
  \epsfig{file=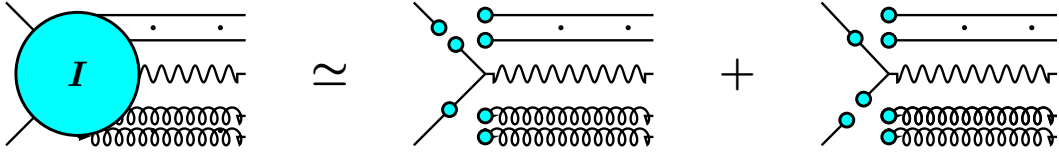,width=14cm}
\caption[dum]{\label{dy_ipt}
  Instanton-induced process for $n_f=3$, 
  $ u_{L}+\bar{u}_{L}\rightarrow 
  \gamma^{\ast} 
   + \bar{d}_{R}+{d}_{R}\,
    +\bar{s}_{R}+{s}_{R}\,          
     +n_g\, {g}$, 
  in leading semiclassical approximation. The amplitude   
  involves the products of the appropriate classical fields
  (lines ending at blobs: fermionic zero modes (straight) and instanton gauge fields (curly)) 
  as well as the non-zero mode quark propagator in the instanton
  background (quark line with central blob).} 
\end{center}
\end{figure}

We start with the derivation of the instanton-induced contribution to 
lepton pair production on the parton level. We will concentrate on the case 
with quarks in the initial state. These contributions dominate over 
the ones involving initial state gluons, at least for scattering
processes where valence-like quarks and antiquarks contribute, e.g. 
in $p\bar{p}$ or $\pi^\pm N$ collisions. This is certainly different 
at very high energies where very small parton momentum fractions $x$ dominate. 
Since the lower bound on $x$ is set by $M^2/S$,
where $M^2$ is the invariant mass squared of the lepton pair and $S$ is 
the hadron-hadron center of mass energy squared, 
the contributing values of $x$ considered in our study are not so small for our 
chosen values of $M^2$ and $S$, see Section \ref{hadron}.   
 For the main case of phenomenological interest, i.e. 
$n_f=3$ light flavours ($m_q\langle\rho\rangle \ll 1$, for $q=u,d,s$),  
instanton-induced quark anti-quark annihilation involves in the final state  
at least two quarks and two anti-quarks of different flavour, such that 
the chirality is violated by $2n_f=6$~\cite{'tHooft:1976up},
plus an arbitrary number of gluons ($g$), e.g. (cf. Fig.~\ref{dy_ipt})  
\begin{eqnarray}
  u_{L}+\bar{u}_{L}&\rightarrow & 
  \gamma^{\ast}
   + \bar{d}_{R}+{d}_{R}\,
    +\bar{s}_{R}+{s}_{R}\,          
     +n_g\, {g}\,.
     \\[.3ex] \nonumber
&& \hspace{.25ex} \hookrightarrow \ell^++\ell^-\
\end{eqnarray}

The amplitudes for the related processes in deep inelastic scattering,
\begin{equation}
  \gamma^{\ast}+g\rightarrow 
   \bar{u}_{R}+{u}_{R}\, 
   + \bar{d}_{R}+d_{R}\,
    +\bar{s}_{R}+s_{R}\,          
     +(n_g-1)\, g,
\end{equation}
have been derived, in leading-order semiclassical approximation, in Ref.~\cite{Moch:1996bs}. 
For clarity and simplicity, let us concentrate here on the explicit result for the 
simplest appropriate\footnote{Note that the even simpler case $n_g=0$ is not relevant for us, since the
corresponding process in lepton pair production would contribute only at vanishingly small 
transverse momentum where, anyhow, instanton perturbation theory is not
applicable (see below).} 
case $n_f=n_g=1$ (cf. Fig.~\ref{dis_ipt_simple}~(left)), 
\begin{eqnarray}
\label{ampifin}
&& {\mathcal T}_{\mu \,\mu ^{\prime}}^a\,
  \left( \gamma^\ast(q) + {g}(p)\rightarrow
    \bar {q}_R(k_1) + {q}_R(k_2) \right)   
   = 
\\[2.4ex]
\nonumber
&& -{\rm i}\,e_q\,\lambda^a \frac{\sqrt{2}}{8}\pi^{3}\,d\, 
     \left( \frac{2\,\pi}{\alpha_s(\mu _{r} )}\right)^{13/2}
    {\exp}\left[-\frac{2\,\pi}{\alpha_s(\mu _{r} )}\right]\,
   2^{\,b}\ \Gamma \left(\frac{b+1}{2}\right)
   \Gamma \left(\frac{b+3}{2}\right)
\\[2.4ex]
&&
\times \, 
\chi_R^\dagger (k_2)
\left[ \left(
\sigma_{\mu ^{\prime}}\overline{p}-p\overline{\sigma}_{\mu ^\prime}\right)
v(q,k_1;\mu _{r} )\overline{\sigma}_\mu 
-\sigma_\mu \overline{v}(q,k_2;\mu _{r} )
\left(
\sigma_{\mu ^\prime}\overline{p} - p\overline{\sigma}_{\mu ^\prime}
\right)
\right]
\chi_L(k_1) ,
\nonumber
\end{eqnarray}
with the four-vector $v_{\lambda}$,
\begin{eqnarray}
  \label{v}  
&&v_\lambda (q,k;\mu _{r} ) \equiv 
\\[1.6ex]
\nonumber
&& 
\frac{1}{\mu _{r}}\,\left\{
  \left[  
\frac{\left( q-k\right)_\lambda}{-(q-k)^2}
+\frac{k_{\lambda}}{2 q\cdot k}
\right] \left(
\frac{\mu _{r}^2}{ -\left( q-k\right)^2}\right)^{\frac{b+1}{2}}
-
\frac{k_{\lambda}}{2 q\cdot k}
\left( \frac{\mu _{r}^2}{-q^{2}}\right)^{\frac{b+1}{2}} 
\right\} \,, 
\end{eqnarray}
and confront it with its chirality conserving counterpart from ordinary perturbation theory 
(cf. Fig.~\ref{dis_ipt_simple} (right)),
\begin{eqnarray}
\label{ampl_pt}
\lefteqn{{\mathcal T}^{a}_{\mu \,\mu ^{\prime}}
     \left( \gamma^{\ast}(q)+{g}(p)\rightarrow
    \bar {q}_{R}(k_1)+{q}_{L}(k_2)\right) =} 
\\[0.5ex]
\nonumber
&& 
e_{q}\,g_s\,\frac{\lambda^{a}}{2}\,  \chi_{L}^{\dagger}(k_{2}) \left[
    \overline{\sigma}_{\mu ^{\prime}}\frac{(q-k_{1})}{(q-k_{1})^{2}}
     \overline{\sigma}_\mu 
       -\overline{\sigma}_\mu \frac{(q-k_{2})}{(q-k_{2})^{2}}
      \overline{\sigma}_{\mu ^{\prime}} \right]
  \chi_{L}(k_{1})\ .
\end{eqnarray}

\begin{figure}
\begin{center}
  \epsfig{file=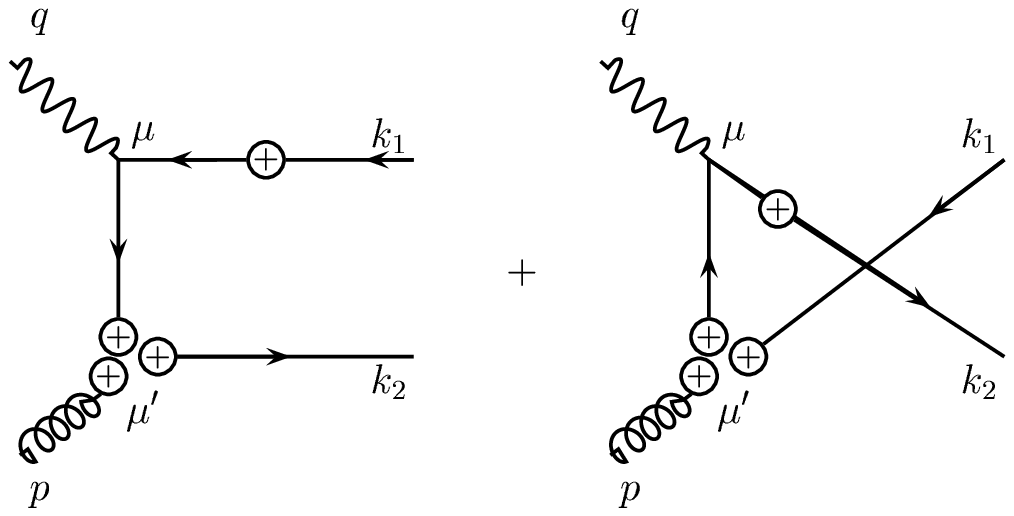,width=7.8cm}
\hspace{7.6ex}
  \epsfig{file=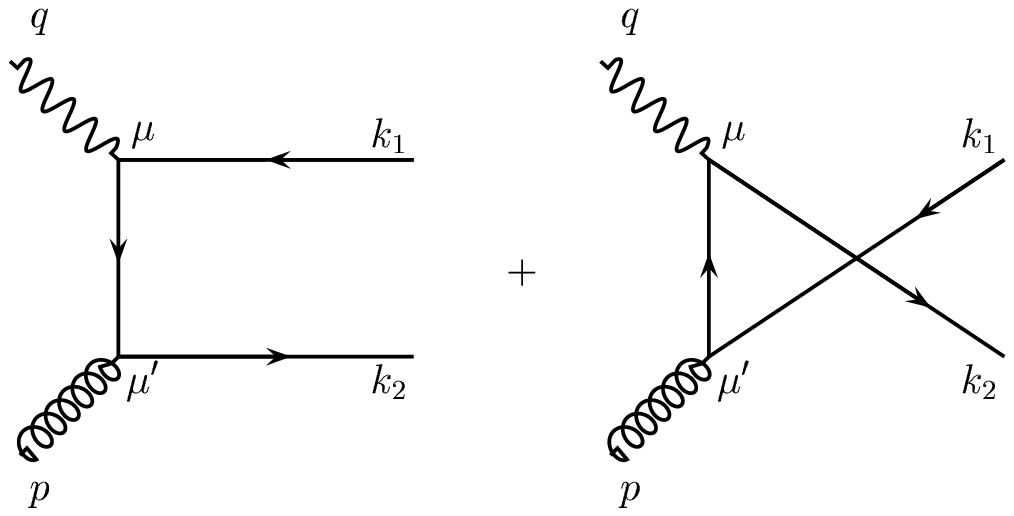,width=7.8cm}
\caption[dum]{\label{dis_ipt_simple}
  Instanton-induced process for $n_f=n_g=1$, $\gamma^{\ast}(q)
  +g(p )\rightarrow \bar q_{R}(k_{1})+{q}_{R}(k_{2})$, 
  in leading semiclassical approximation (left) 
  and the analogous process, $\gamma^{\ast}(q)
  +g(p )\rightarrow \bar q_{R}(k_{1})+{q}_{L}(k_{2})$, 
  from ordinary perturbation theory (right). Both figures from Ref.~\cite{Moch:1996bs}.}
\end{center}
\end{figure}

Here, $e_q$ is the quark charge in units of the electric charge $e$, $g_s$ is the 
strong coupling, 
$\lambda^a$, $a=1,...,8$, are the Gell-Mann $SU(3)$ generators, and $\mu$ and $\mu^\prime$ 
are the four-vector indices of the photon and gluon, respectively
(cf. Fig.~\ref{dis_ipt_simple}). 
The two-component Weyl-spinors $\chi_{L,\,R}$ in Eqs.~(\ref{ampifin}) and (\ref{ampl_pt}) satisfy
the Weyl-equations, 
$\overline{k}\,\chi_L(k)=0$, $k\,\chi_R(k) =0$, and 
the relations $\chi_L(k)\,\chi_L^\dagger (k) = k$, 
$\chi_R(k)\,\chi_R^\dagger (k) = \overline{k}$.
We used the abbreviations,
$k \equiv k_\mu \,\sigma^\mu$, $\overline{k} \equiv k_\mu\,\overline{\sigma}^\mu$, 
for any four-vector $k_\mu $,
with the familiar $\sigma$-matrices, $\sigma_\mu =(1,\vec{\sigma})$ and 
$\overline{\sigma}_\mu =(1,-\vec{\sigma})$, with $\vec{\sigma}$ being the Pauli matrices.

The other parameters in Eq.~(\ref{ampifin}) arose from (the integration over)
the instanton size distribution~\cite{'tHooft:1976up,Bernard:1979qt}, 
whose two-loop renormalization group improved form~\cite{Morris:1984zi}, 
\begin{equation}
  D(\rho ,\mu _{r} )= \frac{d}{\rho^5}\,\left( \frac{2\,\pi}{\alpha_s(\mu _{r} )}\right)^6\, 
     {\exp}\left[-\frac{2\,\pi}{\alpha_s(\mu _{r} )}\right] \,
     \left(
    \rho\,\mu _{r} \right)^{\beta_0+\frac{\alpha_s(\mu _{r} )}{4\,\pi}\, 
     \left( \beta_1 - 12\,\beta_0 \right)}\, ,
\label{density}
\end{equation}
has been exploited, where $\mu_r$ is the renormalization scale and 
\begin{equation}
\label{alpha}
\alpha_{s}(\mu _{r})\equiv \frac{g_s^2(\mu_r )}{4\pi} = 
\frac{4\,\pi}{\beta_{0}\,\ln \left(\frac{\mu _{r}^{2}}{\Lambda^{2}}\right)}
\left[ 1 - \frac{\beta_{1}}{\beta_{0}^{2}}
\frac{\ln\left(\ln \left(\frac{\mu _{r}^{2}}{\Lambda^{2}}\right) \right)}
{\ln \left(\frac{\mu _{r}^{2}}{\Lambda^{2}}\right)}
\right] 
\end{equation}
is the strong fine structure constant at two-loop, with
\begin{equation}
\label{beta}
  \beta_0=11-\frac{2}{3}\,n_f\, ,\hspace{6ex} 
  \beta_1=102-\frac{38}{3}\,n_f 
\end{equation}
being the familiar perturbative coefficients of the QCD beta-function.
The constant $d$ is given by 
\begin{equation}
\label{d}
  d=\frac{C_1}{2}\,{\rm e}^{-3\,C_2+n_f\,C_3}\, ,
\end{equation}
with $C_{1}=0.466$, $C_{2}=1.51$, and $C_{3}=0.292$, in the 
$\overline{\rm MS}$-scheme~\cite{Hasenfratz:1981tw,Luscher:1981zf,'tHooft:1986nc}. 
The variable $b$ in Eq.~(\ref{ampifin}) is a shorthand for the
effective power of $\rho\mu _{r}$ in the instanton size distribution~(\ref{density}),
\begin{equation}
\label{beff}
b\equiv
\beta_0+\frac{\alpha_s(\mu _{r} )}{4\,\pi}\, \left( \beta_1 -
      12\,\beta_0 \right) \, .
\end{equation}
It is important to note that the perturbative
expression~(\ref{density}) for the size distribution is valid for
small $\rho\Lambda\ll 1$ where $\Lambda$ is the fundamental scale in
QCD. Indeed, a comparison with lattice data from quenched ($n_f=0$)
QCD~\cite{Smith:1998wt} yields $\rho\Lambda\lwig 0.4$ for the fiducial
region of instanton perturbation theory~\cite{Ringwald:1999ze}. This
can be translated into a fiducial kinematical region for instanton
perturbation theory in deep inelastic scattering.  On account of the
fact that the main contribution to the integration over the instanton
size comes from~\cite{Moch:1996bs}
\begin{equation}
\langle \rho\rangle \simeq \frac{b+3/2}{\sqrt{-q^2}}\,, \frac{b+3/2}{\sqrt{-(q-k_1)^2}}
\,, \frac{b+3/2}{\sqrt{-(q-k_2)^2}}\,,
\end{equation}
corresponding to different terms in Eqs.~(\ref{ampifin}) and (\ref{v}),  
one has to require that all virtualities, 
$\sqrt{-q^2}$, $\sqrt{-(q-k_1)^2}$, and $\sqrt{-(q-k_2)^2}$, 
exceed $\mathcal Q_{\rm min}\approx (4-6)$~GeV, in order to stay in the realm of instanton 
perturbation theory.  

\begin{figure}
\begin{center}
  \epsfig{file=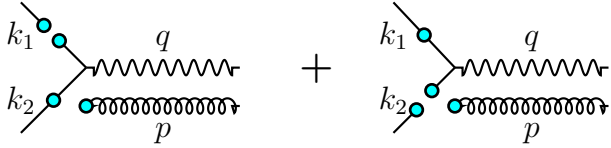,width=8cm}
\caption[dum]{\label{lpp_ipt_simple}
  Instanton-induced process for $n_f=n_g=1$, 
  $q_{L}(k_{1})+\bar {q}_{L}(k_{2})\rightarrow\gamma^{\ast}(q) +g(p ) $, 
  in leading semiclassical approximation.}
\end{center}
\end{figure}

It is now straightforward to obtain the corresponding amplitudes relevant for lepton pair production 
via quark anti-quark annihilation, namely the one for the chirality violating instanton-induced process
$q_{L}(k_{1})+\bar {q}_{L}(k_{2})\rightarrow\gamma^{\ast}(q) +g(p )$ 
(cf. Fig.~\ref{lpp_ipt_simple}) 
and the one for the analogous ordinary perturbative process 
$q_{L}(k_{1})+\bar {q}_{R}(k_{2})\rightarrow\gamma^{\ast}(q) +g(p )$. 
In fact, these processes are basically $T$-conjugates of the deep inelastic
processes from Eqs.~(\ref{ampifin}) and (\ref{ampl_pt}), and the respective modulus-squared amplitudes 
are therefore identical, up to reflections of three-momenta. 
Some care has of course to be taken with respect to the photon
virtuality: whereas in deep inelastic scattering it is space-like, $Q^2\equiv -q^2>0$, 
in lepton pair production it is time-like, $M^2\equiv q^2>0$. We will comment on this later.  

We have calculated the contribution of our simple processes to the partonic 
tensor\footnote{Averaging over colour and spin of the initial state is implicitly understood
in Eq.~(\ref{wmunu}); the index $n$ is to label besides the final
state partons also their spin and colour degrees of freedom.}
for inclusive quark anti-quark annihilation into a virtual photon, 
\begin{eqnarray}
  {w}_{\mu \nu} (k_1,k_2;q ) & =&  \sum_{n=0}^\infty
  {w}_{\mu \nu}^{(n)} (k_1,k_2;q ) \, ,
\label{wmunu}
\\[10pt] 
 {w}_{\mu \nu}^{(n)} (k_1,k_2;q )& =& 
\frac{1}{4\,\pi}\,
  \int {\rm dPS}^{(n)}\, 
    {\mathcal T}_\mu (k_1,k_2;q,p_{1},\ldots,p_{n}) 
    {\mathcal T}_{\nu}^\ast(k_1,k_2;q,p_{1},\ldots ,p_{n} )  ,
\label{wmunu_n}
\\[10pt]
\int {\rm dPS}^{(n)} &=& 
\prod_{j=1}^{n} \int
\frac{{\rm d}^{4}p_{j}}{(2\,\pi)^{3}}\, \delta^{(+)}\left( p_{j}^{2}\right)
(2\,\pi)^{4}\, \delta^{(4)}\left( k_1+k_2-q-p_{1}-\ldots -p_{n}\right) \, .
\end{eqnarray}
Following Ref.~\cite{Lam:1980uc}, this tensor can be decomposed as
\begin{eqnarray}
w_{\mu\nu}=-\tilde{g}_{\mu\nu}\*w_1+\tilde{K}_{\mu}^{\prime}\*\tilde{K}_{\nu}^{\prime}\*w_2
-{\tilde{K}_{\mu}^{\prime}\tilde{k}_{\nu}^{\prime}
+\tilde{K}_{\nu}^{\prime}\tilde{k}_{\mu}^{\prime}\over 2}\*w_3
+\tilde{k}_{\mu}^{\prime}\*\tilde{k}_{\nu}^{\prime}\*w_4,
\label{wmunu_k}
\end{eqnarray}
with 
$K^{\prime}=k_1+k_2$, $k^{\prime}=k_1-k_2$, 
$\tilde{g}_{\mu\nu}=g_{\mu\nu}-{q_\mu\*q_{\nu}/ q^2}$, 
$\tilde{K}_\mu^{\prime} = {\tilde{g}_{\mu\nu}\*K^{\prime\,\nu}/ \sqrt{s}}$, and
$\tilde{k} _\mu^{\prime}= {\tilde{g}_{\mu\nu}\*k^{\prime\,\nu}/ \sqrt{s}}$, where
$s=(k_1+k_2)^2$.
In turn, the different $w_i$ can be obtained,  
\begin{equation}
w_i = p^i_{\mu\nu}\*w^{\mu\nu},\qquad i=1,2,3,4\,,
\end{equation}
from the partonic tensor with the help of the following projectors, 
\begin{eqnarray}
\label{p0}
p^0_{\mu\nu}&=&-g_{\mu\nu}, \\
p^1_{\mu\nu}&=&{4\over s\*t\*u}\*
\varepsilon_{\mu\rho\sigma\tau}\*k_1^{\rho}\*k_2^{\sigma}\*q^{\tau}\*
\varepsilon_{\nu\rho'\sigma'\tau'}
\*k_1^{\rho'}\*k_2^{\sigma'}\*q^{\tau'},  \\
p^2_{\mu\nu}&=&  {-q^2\*s\over t\*u}\*\left[
\tilde{k}^{\prime\,2}\*\left(p^0_{\mu\nu}-2\*p^1_{\mu\nu}\right)
+\tilde{k}^{\prime}_{\mu}\*\tilde{k}^{\prime}_{\nu}
\right], \\
p^3_{\mu\nu}&=&  {-2\*q^2\*s\over t\*u}\*\left[
(\tilde{k}^{\prime}\cdot\tilde{K}^{\prime})\*\left(p^0_{\mu\nu}-2\*p^1_{\mu\nu}\right)
+{\tilde{k}_{\mu}^{\prime}\*\tilde{K}_{\nu}^{\prime}+\tilde{K}_{\mu}^{\prime}\*\tilde{k}_{\nu}^{\prime}\over 2}
\right], \\
\label{p4}
p^4_{\mu\nu}&=&  {-q^2\*s\over t\*u}\*\left[
\tilde{K}^{\prime\,2}\*\left(p^0_{\mu\nu}-2\*p^1_{\mu\nu}\right)
+\tilde{K}_{\mu}^{\prime}\*\tilde{K}_{\nu}^{\prime}
\right],  
\end{eqnarray}
where $t=(q-k_1)^2$ and $u=(q-k_2)^2$.  
We are especially interested in the contribution $w_{\mu\nu}^{(1)}$ to
$w_{\mu\nu}$ (\ref{wmunu}),
\begin{eqnarray}
   w^\prime_{\mu\nu}&=& {\overline{\sum}_{\rm\, spins,colours}}
 \,{\mathcal T}_\mu (k_1,k_2;q,p) 
    {\mathcal T}_{\nu}^\ast(k_1,k_2;q,p) \,,\qquad
w^\prime_i=p^i_{\mu\nu}   w^{\prime\,\mu\nu} \,,
\\
 w_i^{(1)}&=&
\frac{1}{4\pi}\int\frac{{\rm d}^4p}{(2\pi)^3}\delta^{(+)}(p^2)\,(2\pi)^4\,\delta^{(4)}(k_1+k_2-q-p)\,w^\prime_i
=\frac{1}{2}\delta^{(+)}((k_1+k_2-q)^2)w^\prime_i\,.
\end{eqnarray}
Along these lines, we find that the simple instanton-induced process 
$q_L+\bar q_L\to \gamma^\ast + g$ contributes as follows to the functions
$w^\prime_i$,
\begin{eqnarray}
\label{wsigma}
-g^{\mu\nu}\*w_{\mu\nu}^{({\rm I})\prime}&=&\xi^{\rm (I)}\*
\Bigg\{
\left({M^2\over -t}\right)^{b+1}+\left({M^2\over -u}\right)^{b+1} 
+{2\*t\*u\over (t-M^2)\*(u-M^2)}\* \\ &\times&
\left[
\left({M^2\over -t}\right)^{b+1\over 2}
\*\left({M^2\over -u}\right)^{b+1\over 2}
+1-{\rm Re}\left(-1\right)^{b+1\over 2}
\*\left(\left({M^2\over -t}\right)^{b+1\over 2}
+\left({M^2\over -u}\right)^{b+1\over 2}\right)
\right]
\Bigg\}, \nonumber \\
\label{w1}
w_1^{({\rm I})\prime}&=&{\xi^{\rm (I)}\over 2}\*\left\{
\left({M^2\over -t}\right)^{b+1\over 2}+\left({M^2\over -u}\right)^{b+1\over 2}
\right\}^2 ,
 \end{eqnarray}
 \begin{eqnarray}
w_2^{({\rm I})\prime}&=&{\xi^{\rm (I)}\over 2}\*\Bigg\{
-{s\*M^2\over (t-M^2)^2}\*\left({M^2\over -t}\right)^{b+1}
-{s\*M^2\over (u-M^2)^2}\*\left({M^2\over -u}\right)^{b+1}
+{t\*u\*(t-u)^2\over (t-M^2)^2\*(u-M^2)^2} \nonumber \\
&-&{s\*M^2\*(4\*M^2\*s+t^2+u^2)\over t\*u\*(t-M^2)\*(u-M^2)}
\*\left({M^2\over -t}\right)^{{b+1\over 2}}\*
\left({M^2\over -u}\right)^{{b+1\over 2}}
+ {(t-u)\*{\rm Re}\left(-1\right)^{b+1\over 2}\over 
(t-M^2)^2\*(u-M^2)^2}\nonumber \\
&\times& 
\Big[
(t-M^2)\*(2\*M^4-M^2\*t-t\*u-3\*M^2\*u+u^2)
\*\left({M^2\over -u}\right)^{{b+1\over 2}}
\nonumber \\
&-&
(u-M^2)\*(2\*M^4-M^2\*u-t\*u-3\*M^2\*t+t^2)
\*\left({M^2\over -t}\right)^{{b+1\over 2}}
\Big]
 \Bigg\} ,
\\
w_3^{({\rm I})\prime}&=&\xi^{\rm (I)}\*
\Bigg\{
{s\*M^2\over (t-M^2)^2}\*\left({M^2\over -t}\right)^{b+1}
-{s\*M^2\over (u-M^2)^2}\*\left({M^2\over -u}\right)^{b+1}
-{(s+M^2)\*(t-u)\*t\*u\over (t-M^2)^2\*(u-M^2)^2}\nonumber \\
&+& {s\*M^2\*(s+M^2)\*(t-u)\over (t-M^2)\*(u-M^2)\*t\*u}
\*\left({M^2\over -t}\right)^{{b+1\over 2}}\*
\left({M^2\over -u}\right)^{{b+1\over 2}}
-{{\rm Re}\left(-1\right)^{b+1\over 2}\over 
(t-M^2)^2\*(u-M^2)^2} \\
&\times& \Big[
(t-M^2)\*(2\*M^6-2\*M^4\*t-4\*M^4\*u+M^2\*t^2+3\*M^2\*u^2+t^2\*u-u^3)
\*\left({M^2\over -u}\right)^{{b+1\over 2}}
\nonumber \\
&-&
(u-M^2)\*(2\*M^6-2\*M^4\*u-4\*M^4\*t+M^2\*u^2+3\*M^2\*t^2+t\*u^2-t^3)
\*\left({M^2\over -t}\right)^{{b+1\over 2}}
\Big]
\Bigg\} ,
\nonumber \\
w_4^{({\rm I})\prime}&=&{\xi^{\rm (I)}\over 2}\*\Bigg\{
-{s\*M^2\over (t-M^2)^2}\*\left({M^2\over -t}\right)^{b+1}
-{s\*M^2\over (u-M^2)^2}\*\left({M^2\over -u}\right)^{b+1}
+{t\*u\*(s+M^2)^2\over (t-M^2)^2\*(u-M^2)^2} \nonumber \\
&-&{s\*M^2\*(t^2+u^2)\over  t\*u\*(t-M^2)\*(u-M^2)}
\*\left({M^2\over -t}\right)^{{b+1\over 2}}\*
\left({M^2\over -u}\right)^{{b+1\over 2}}
+{(s+M^2)\*{\rm Re}\left(-1\right)^{b+1\over 2}\over 
(t-M^2)^2\*(u-M^2)^2}\nonumber \\
&\times& \Big[(t-M^2)\*(-M^2\*u+M^2\*t+t\*u+u^2)
\*\left({M^2\over -u}\right)^{{b+1\over 2}}
 \nonumber \\
&+&(u-M^2)\*(-M^2\*t+M^2\*u+t\*u+t^2)
\*\left({M^2\over -t}\right)^{{b+1\over 2}}
\Big]
\Bigg\}
\label{w4}
 ,
\end{eqnarray}
where  
\begin{eqnarray}
\label{xieff_i}
\xi^{\rm (I)} &\equiv & \pi^2\*e_q^2\*{\cal N}^2\*\left(
{2\*\pi\over \alpha_s(\mu_r)}\right)^{13}\*\exp\left(-{4\*\pi\over \alpha_s(\mu_r)}\right)
\*\left({\mu_r^2\over M^2}\right)^b\*{s\over M^2},
\\
{\mathcal N} &\equiv &\frac{1}{2}\,\pi^{2}\,d\ 
     2^{\,b}\,  \Gamma \left(\frac{b+1}{2}\right)\,
   \Gamma \left(\frac{b+3}{2}\right)\, .
\end{eqnarray}
We have obtained these results starting from Eq.~(\ref{ampifin}), 
contracting it with the gluon polarization vector $\epsilon^{\mu^\prime}(p)$ 
and taking the modulus squared, exploiting FORM~\cite{Vermaseren:2000nd} for the spinor traces.   
The results~(\ref{wsigma})--(\ref{w4}) for the contribution of our simple 
instanton-induced process to the partonic tensor 
for inclusive quark anti-quark annihilation into a time-like photon look quite similar to 
the contribution of the analogous simple instanton-induced process to the deep inelastic 
structure tensor of a gluon found in Ref.~\cite{Moch:1996bs}. 
A notable difference is the appearance of the 
factor ${\rm Re} (-1)^{(b+1)/2}$, which reduces to unity in the space-like kinematics
of deep inelastic scattering and was therefore not visible in the results of 
Ref.~\cite{Moch:1996bs}. On the other hand, the full instanton contribution to the 
deep inelastic structure tensor of a gluon can be obtained from Eqs.~(\ref{wsigma})--(\ref{w4}) 
by replacing the combinatorial factor $1/((2\cdot N_c)^2)$ by $1/(2\cdot (N_c^2-1))$ and by substituting 
${\rm Re} (-1)^{(b+1)/2}$ by $1$.     

As a check, let us also quote the corresponding perturbative contributions
to the inclusive partonic tensor arising from Eq.~(\ref{ampl_pt}),
\begin{eqnarray}
\label{w0_pt}
-g^{\mu\nu}\*w_{\mu\nu}^{({\rm pt})\prime}&=&\xi^{\rm (pt)}\*\left[
{u\over t}+ {t\over u} + {2\*M^2\*s\over t\*u}
\right], \\
\label{w1_pt}
w_1^{({\rm pt})\prime}&=& 
-{1\over 2}\*g^{\mu\nu}\*w_{\mu\nu}^{({\rm pt})\prime},
 \\
\label{w2_pt}
w_2^{({\rm pt})\prime}&=& -\xi^{\rm (pt)}\*{s\*M^2\over t\*u},
 \\
\label{w3_pt}
w_3^{({\rm pt})\prime}&=& 0,
 \\
\label{w4_pt}
w_4^{({\rm pt})\prime}&=& w_2^{({\rm pt})\prime},
\end{eqnarray}
where
\begin{equation}
\label{xieff_pt}
\xi^{\rm (pt)} \equiv 2\pi e_q^2\*\alpha_s.
\end{equation}
We obtained this well known result, see e.g. \cite{Lam:1978zr}, by
exploiting the same FORM routines as the ones for the
instanton-induced contribution, except for replacing the input
amplitude~(\ref{ampifin}) by the perturbative
amplitude~(\ref{ampl_pt}).

\section{\label{angular}The angular distribution of the lepton pairs} 

In this Section, we will concentrate on the angular distribution of lepton pairs
in instanton-induced processes -- mainly concentrating on the simple one from 
the previous Section -- and compare it to the one predicted from 
ordinary perturbation theory.  

In general, the angular distribution of the charged lepton $\ell^+$ in lepton pair production, 
\begin{eqnarray}
\label{drellyan_had}
h_1(K_1)+h_2(K_2)&\rightarrow& \gamma^\ast (q) + X\,,
\\[.3ex] \nonumber &&\hspace{0.1ex} \hookrightarrow \ell^+(q_+) + \ell^- (q_-)
\end{eqnarray}
is described by three 
functions, $\lambda$, $\mu$, and $\nu$, which may depend on the 
kinematic variables of (\ref{drellyan_had}), 
\begin{equation}
\label{ang-dist}
\frac{1}{\sigma}\frac{d\sigma}{d\Omega}=\frac{3}{4\pi}\,\frac{1}{\lambda+3}
\left(1+\lambda\,\cos^2\!\theta
+\mu\,\sin 2\theta\,\cos\phi
+\frac{\nu}{2}\sin^2\!\theta\,\cos 2\phi \right)\,,
\end{equation}
$\theta$ and $\phi$ being the polar and azimuthal angles of $\ell^+$, 
respectively~\cite{Collins:1977iv}. 
These coefficient functions may be conveniently expressed in terms of 
hadronic helicity structure functions~\cite{Lam:1978pu},
\begin{eqnarray}
\label{lambda}
\lambda &=& {W_T-W_L\over W_T+W_L},\\
\label{mu}
\mu &=& \frac{W_\Delta}{W_T+W_L},\\
\label{nu}
\nu &=& {2\,W_{\Delta\Delta}\over W_T+W_L}.
\end{eqnarray}
Here, we exploit the so-called Collins-Soper frame~\cite{Collins:1977iv}, in which the 
frame dependent helicity structure functions read, in terms of the 
hadronic counterparts $W_i$ of the 
previously introduced invariant functions $w_i$,  
\begin{eqnarray}
W_T &=& W_1+{r^2\over 2\*S\*M^2\*(1+r^2)}\*\left[
(q\cdot K)^2\*W_2-(q\cdot K)\*(q\cdot k)\*W_3+(q\cdot k)^2\*W_4\right],\\
 W_L &=& W_1+{1\over S\*M^2\*(1+r^2)}\*\left[
(q\cdot k)^2\*W_2-(q\cdot K)\*(q\cdot k)\*W_3+(q\cdot K)^2\*W_4\right],\\
W_{\Delta} &=& -{r\over S\*M^2\*(1+r^2)}\*\left[
-(q\cdot K)\*(q\cdot k)\*(W_2+W_4)+{(q\cdot K)^2+(q\cdot k)^2\over 2}
\*W_3\right],\\
W_{\Delta\Delta} &=& -{r^2\over 2\*S\*M^2\*(1+r^2)}\*\left[
(q\cdot K)^2\*W_2-(q\cdot K)\*(q\cdot k)\*W_3+(q\cdot k)^2\*W_4\right],
\end{eqnarray}
where
\begin{equation}
r^2\equiv {q_\perp^2\over M^2}={T\*U+M^2(M^2-S-T-U)\over S\*M^2}\,,
\label{R-eq}
\end{equation}
determines the transverse photon momentum $q_\perp$ with respect to the
hadronic reaction plane. The kinematic variables $S=(K_1+K_2)^2$,
$T=(q-K_1)^2$ and $U=(q-K_2)^2$ refer to the hadron level (cf.
Eq.~(\ref{drellyan_had})).  Similarly, $K=K_1+K_2$ and $k=K_1-K_2$.

\subsection{\label{partonlevel} Parton level}

The contribution of our simple instanton-induced process as well as the
contributions from ordinary perturbation theory to these helicity structure
functions are determined by folding their partonic counterparts with the 
parton density distributions. Before doing that it is instructive 
to consider first the partonic analogies of the quantities. 

Let us start with the contributions arising from ordinary perturbation theory, 
(cf. Eqs.~(\ref{w0_pt})--(\ref{w4_pt})), 
\begin{eqnarray}
\label{hatlambda_pt}
\hat\lambda^{({\rm pt})(1)} &\equiv & {w_T^{({\rm pt})(1)}-w_L^{({\rm pt})(1)}\over 
w_T^{({\rm pt})(1)}+w_L^{({\rm pt})(1)}}
= {2-r^2\over 2+3\,\*r^2},\\
\hat\mu^{({\rm pt})(1)} &\equiv & \frac{w_\Delta^{({\rm pt})(1)}}
{w_T^{({\rm pt})(1)}+w_L^{({\rm pt})(1)}}=
\frac{q_\perp}{M}
\frac{2s (s+M^2) M^2 (t-u)}{(t-M^2)^2+(u-M^2)^2}
\frac{1}{2M^2s+3tu},\\
\label{hatnu_pt}
\hat\nu^{({\rm pt})(1)} &\equiv & {2\,w_{\Delta\Delta}^{({\rm pt})(1)}\over 
w_T^{({\rm pt})(1)}+w_L^{({\rm pt})(1)}} 
={2\,\*r^2\over 2+3\,\*r^2}.
\end{eqnarray} 
In terms of the partonic quantities, the ratio $r^2=q_\perp^2/M^2$ in Eq.~(\ref{R-eq}) reduces to  
$tu/(sM^2)$. In particular, we find the Lam-Tung
relation~\cite{Lam:1980uc},
\begin{equation}
1-\hat\lambda^{({\rm pt})(1)}-2\,\hat\nu^{({\rm pt})(1)} = 0\,,
\end{equation} 
which, of course, holds also on the hadron level, 
$1-\lambda^{({\rm pt})}-2\,\nu^{({\rm pt})}=0$, as long as no
intrinsic transverse momentum for the initial state quarks is invoked. 
It is nearly left intact even if one includes $\mathcal O(\alpha_s^2)$
corrections~\cite{Mirkes:1994dp}.   

The contributions arising from our simple instanton-induced process 
(cf. Eqs.~(\ref{wsigma})--(\ref{w4})) are readily calculated along the same
lines. They yield quite lengthy expressions, and we do not quote them all
analytically, but will illustrate them, instead, graphically.  
We stress, however, that the Lam-Tung relation is violated by instantons.
This is apparent from the following, non-vanishing expression, 
\begin{eqnarray}
2\left(w_L^{({\rm I})\prime}-2\,w_{\Delta\Delta}^{({\rm I})\prime}\right)
&=&{4\*\xi^{\rm (I)}\*t\*u\over (t-M^2)\*(u-M^2)}\Bigg\{
1-{\rm Re}\left(-1\right)^{b+1\over 2}
\*\left[\left({M^2\over -t}\right)^{b+1\over 2}
+\left({M^2\over -u}\right)^{b+1\over 2}\right]\nonumber \\
&-&{s\*M^2\over t\*u}\*\left({M^2\over -t}\right)^{b+1\over 2}
\*\left({M^2\over -u}\right)^{b+1\over 2}
\Bigg\}
\,,
\label{num}
\end{eqnarray}   
for the numerator of the Lam-Tung combination (cf. Eqs.~(\ref{lambda})--(\ref{nu})), 
\begin{equation}
1-\hat\lambda-2\,\hat\nu = \frac{2\left( w_L-2 w_{\Delta\Delta} \right)}{w_T+w_L}\,.
\end{equation}
Note, that the factor $(M^2/\sqrt{tu})^{b+1}$ in the asymmetry
(\ref{num}) arises from a non-planar diagram. That is in accordance
with \cite{Boer:2004mv} where the importance of non-planar
interference terms for the violation of the Lam-Tung relation were
discussed.

\begin{figure}
\begin{center}
  \epsfig{file=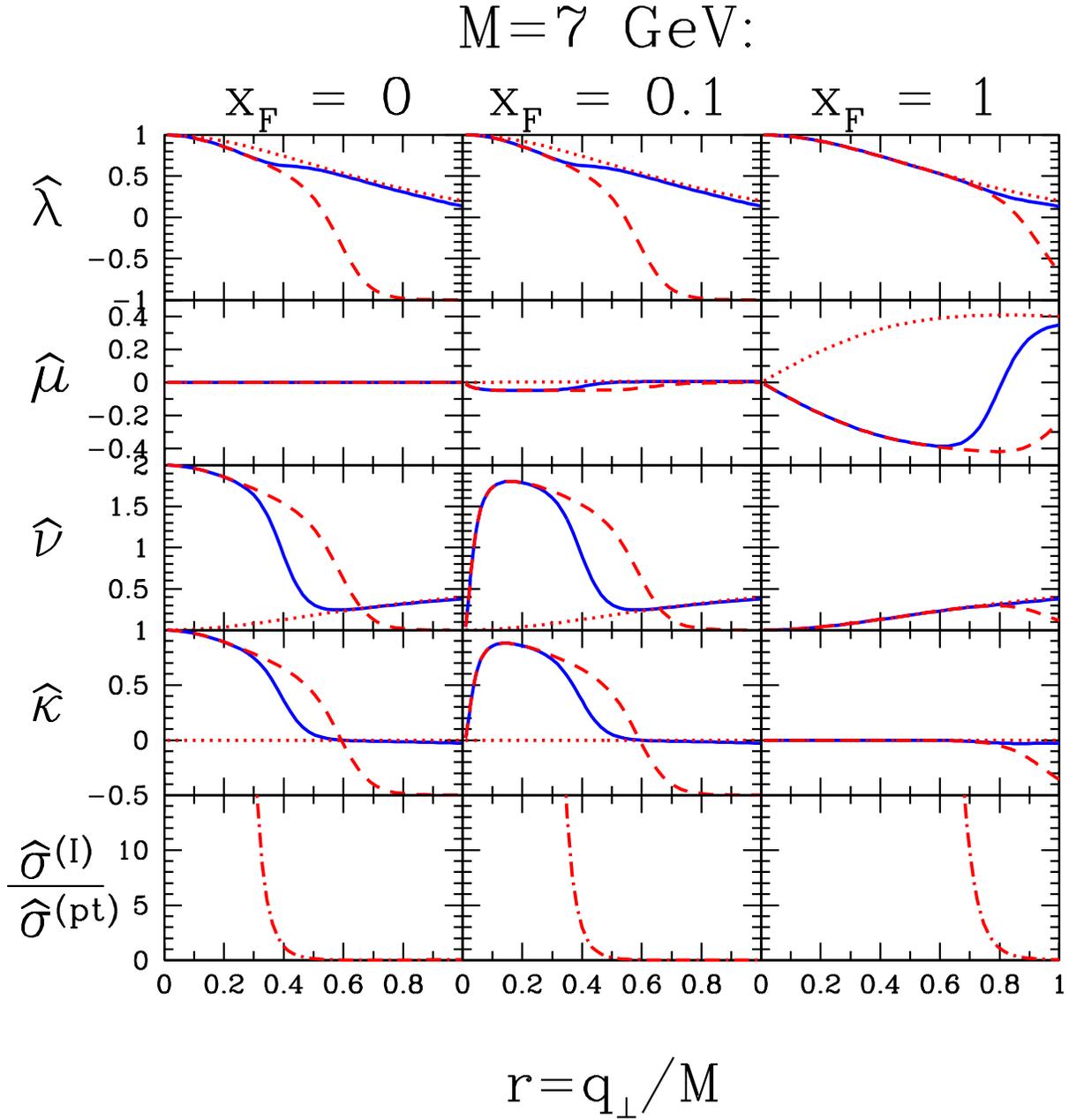,width=17.cm}
\caption[dum]{\label{coeff_funct_part_7}
        Coefficient functions of the angular distribution of lepton pairs 
        from quark anti-quark annihilation,     
        as a function of $q_\perp/M$, for various values of $x_F$, for $M=7$~GeV.
        Dotted: result from pure ordinary perturbation theory; dashed: result from pure instanton
        perturbation theory; solid: total result from ordinary and instanton perturbation theory.
        Also shown, in the last row, is the ratio of the cross-sections of instanton perturbation 
        theory to ordinary perturbation theory (dashed-dotted).
        In the numerical results shown we have chosen $\mu_r=M$, $\Lambda=0.346$~GeV, and $n_f=3$.}
\end{center}
\end{figure}
\begin{figure}
\begin{center}
  \epsfig{file=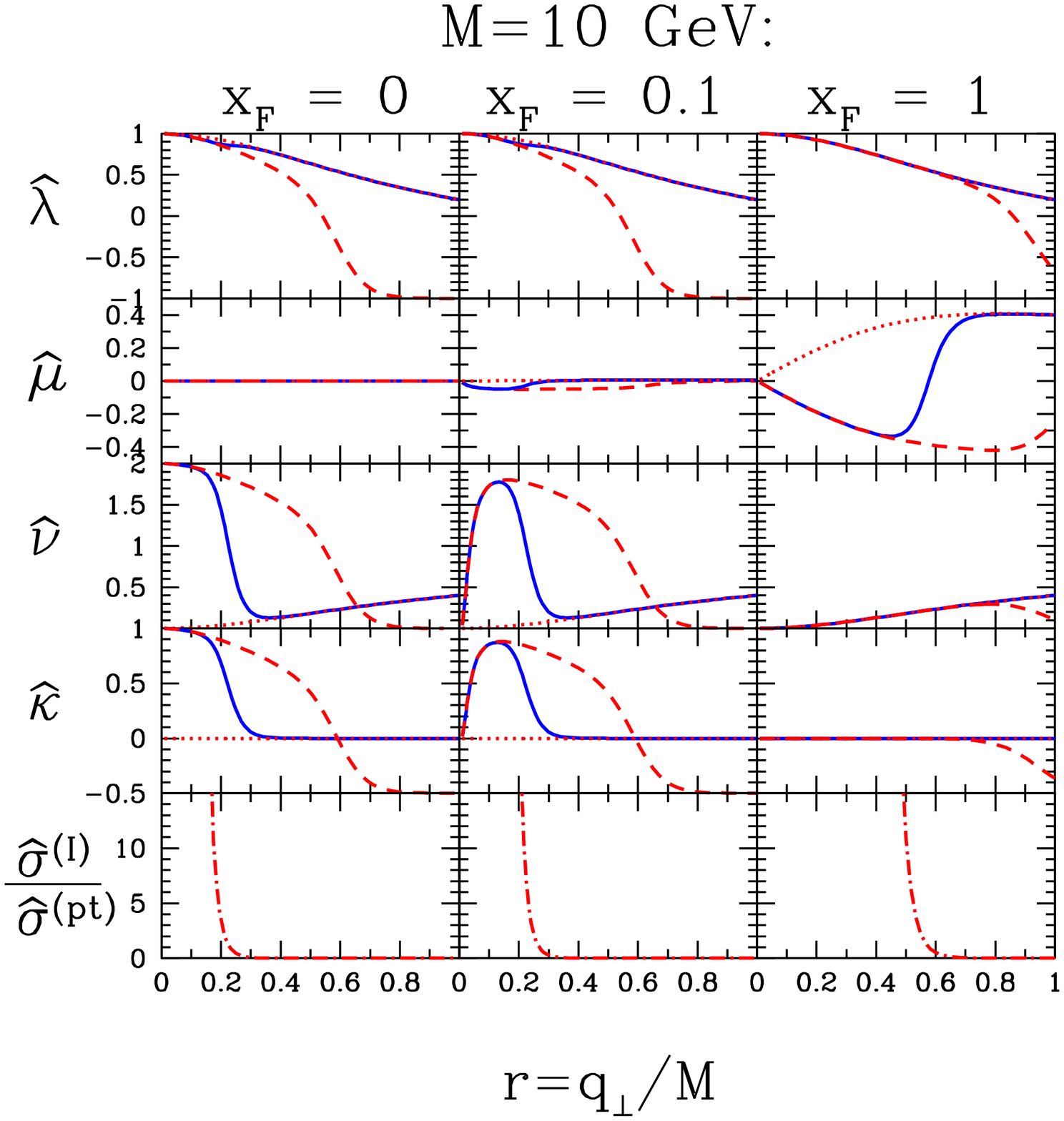,width=17.cm}
\caption[dum]{\label{coeff_funct_part_10}
        Same as in Fig.~\ref{coeff_funct_part_7}, but for $M=10$~GeV.}
\end{center}
\end{figure}

It is useful to view the partonic coefficient functions, for fixed $M$, 
as a function of $r=q_\perp/M$ and the partonic Feynman variable 
\begin{equation}
x_F\equiv \frac{t-u}{s}
\,.
\label{xf_part}
\end{equation}
In fact, $\hat\lambda$, $\hat\mu$, and $\hat\nu$ depend, for fixed $M$, only 
on $r$ and $x_F$. Their dependence on these kinematical variables is illustrated
in Figs.~\ref{coeff_funct_part_7} and \ref{coeff_funct_part_10}. 
We observe the following features:

{(i)} $\hat\lambda$ (top panels) approaches $-1$ for large $r$ for
pure instanton-induced processes (dashed), i.e. these processes tend
to be purely longitudinal (cf. Eq.~(\ref{lambda})) for large
transverse momenta, in contrast to ordinary perturbative processes
(dotted and Eq.~(\ref{hatlambda_pt})).  The total result for
$\hat\lambda$ (solid), taking into account both instanton and ordinary
processes in the numerator and denominator of the partonic equivalent
of Eq.~(\ref{lambda}), shows little deviation from ordinary
perturbation theory. Indeed, there are experimental hints for longitudinally
polarized photons in hadron collisions towards larger $x_F$
\cite{Conway:1989fs}. Note that, even if the instanton-induced process
$q\bar{q}\to\gamma^\ast g$ is suppressed, for larger $x_F$, instanton
effects might be relevant for this effect since gluon resummation
leads to an enhancement in this kinematic region, see Section
\ref{gluons}.

{(ii)} The total result (solid) for $\hat\mu$ (second panels from top) shows a quite 
significant deviation from ordinary perturbation theory (dotted) for sizeable $x_F$ 
and intermediate values of $r$.  

{(iii)} $\hat\nu$ (third panels from top) behaves quite differently in pure 
instanton-induced processes (dashed) and ordinary perturbative processes (dotted and Eq.~(\ref{hatnu_pt})). 
In fact, instanton-induced processes have a value of\footnote{In fact, 
it follows from general arguments~\cite{Brandenburg:1993cj,Boer:2004mv} 
that as long as instanton processes dominate over ordinary perturbative processes, one expects 
$\hat\nu^{({\rm I})}\approx 2$.} $\hat\nu^{({\rm I})}\approx 2$ at small, but
finite $r$ and small $x_F$, much larger as ordinary perturbative processes 
($\hat\nu^{({\rm pt})}\ll 1$).  
Also in the total result for $\hat\nu$ (solid) we observe a strong enhancement 
at small, but finite $r$ and small $x_F$ in comparison to ordinary perturbation theory. 
Correspondingly, we find a strong violation of the Lam-Tung relation, which 
we display in the forth panels from top in terms of the parameter
\begin{equation}
\hat\kappa \equiv -\frac{1}{4}\left( 1 - \hat\lambda -2\,\hat\nu\right)
\,.
\end{equation}
Whereas this parameter is identical zero in ordinary perturbative processes (dotted), 
it is about one, at small $r$, $x_F$, for instanton-induced processes, leading
to a drastic violation of the Lam-Tung relation in the total result 
(solid).    

{(iv)} Clearly, instanton effects in the coefficient functions are most visible 
in kinematical regions where the instanton induced cross-section $\hat\sigma^{\rm (I)}$
dominates over the perturbative, $\hat\sigma^{\rm (pt)}$, one. The instanton-induced
features in Figs.~\ref{coeff_funct_part_7} and \ref{coeff_funct_part_10} are indeed located 
where the ratio (bottom panels; dashed-dotted)
\begin{equation}
\frac{\hat\sigma^{\rm (I)}}{\hat\sigma^{\rm (pt)}}=
\frac{2\,w_T^{\rm (I)} + w_L^{\rm (I)}}{2\,w_T^{\rm (pt)} + w_L^{\rm (pt)}}
\end{equation}
becomes large. Obviously, it gets large towards small momentum transfer. 
The dominance of instantons is seen to set in like a ``brick wall''.
This sudden onset occurs practically at the boundary of the fiducial
kinematical region of instanton perturbation theory,
$\sqrt{-t}\approx\mathcal Q_{\rm min}$ or $\sqrt{-u}\approx \mathcal
Q_{\rm min}$, for $M>\mathcal Q_{\rm min}\approx (4-6)$~GeV.
Therefore, the instanton features in the coefficient functions at very
small momentum transfer, to the left of the sudden onset of instanton
dominance in Figs.~\ref{coeff_funct_part_7} and
\ref{coeff_funct_part_10}, lie strictly speaking outside the range of
validity of the semiclassical approximation.  Fortunately, however,
the coefficient functions are {\it ratios} of helicity structure
functions (cf. Eqs.~(\ref{lambda})--(\ref{nu})), and therefore the
main uncertainties coming from the extrapolation of the perturbative
expression of the instanton-size distribution cancel in them.
Therefore, it is expected that our predictions of the coefficient
functions remain also valid at smallish, but non-zero $r$.  For very
small $q_\perp=M r$, namely up to around 1 GeV, the simplest
perturbative and instanton-induced sub-process
$q\bar{q}\to\gamma^\ast$ contributes and may change the angular
distributions.
 
{(v)} One feature of $\hat{\lambda}^{\rm (pt)}$ and $\hat{\nu}^{\rm
  (pt)}$ that translates unchanged even to the hadron level is the scaling behavior: 
they depend only on the ratio $r$. This is not the case when
instanton effects are included since they vanish in comparison
to the perturbative contributions for larger $M^2$.  This is basically
triggered by the ratio $\hat\sigma^{\rm (I)}/\hat\sigma^{\rm (pt)}$
which leads to a $M^2$ dependent weighting of the perturbative and
instanton contribution. In addition, also $\hat{\lambda}^{\rm (I)}$
and $\hat{\nu}^{\rm (I)}$ depend already slightly on $M^2$.

{(vi)} Remarkable is also the behavior for $x_F=0$. For vanishing
$q_\perp$, one expects to recover the well known leading-order angular
distribution $\propto(1+\cos^2\theta)$, that is $\hat{\lambda}= 1$ 
and $\hat{\nu}, \hat{\mu}= 0$.  As one can see from
Fig.~\ref{coeff_funct_part_7} and \ref{coeff_funct_part_10}, the
function $\hat{\nu}$ tends towards 2.  Note that $\hat{\nu}$ still
vanishes in the limit $q_\perp\to 0$ for very small but finite $x_F$.
 For large $x_F$, the violation of the Lam-Tung relation is suppressed
even in a region where the ratio $\hat\sigma^{\rm (I)}/\hat\sigma^{\rm
  (pt)}$ is not small. The very strong $x_F$ dependence
will disappear after folding with the parton distributions as we
will see in the next section.

\begin{figure}[t]
\begin{center}
  \epsfig{file=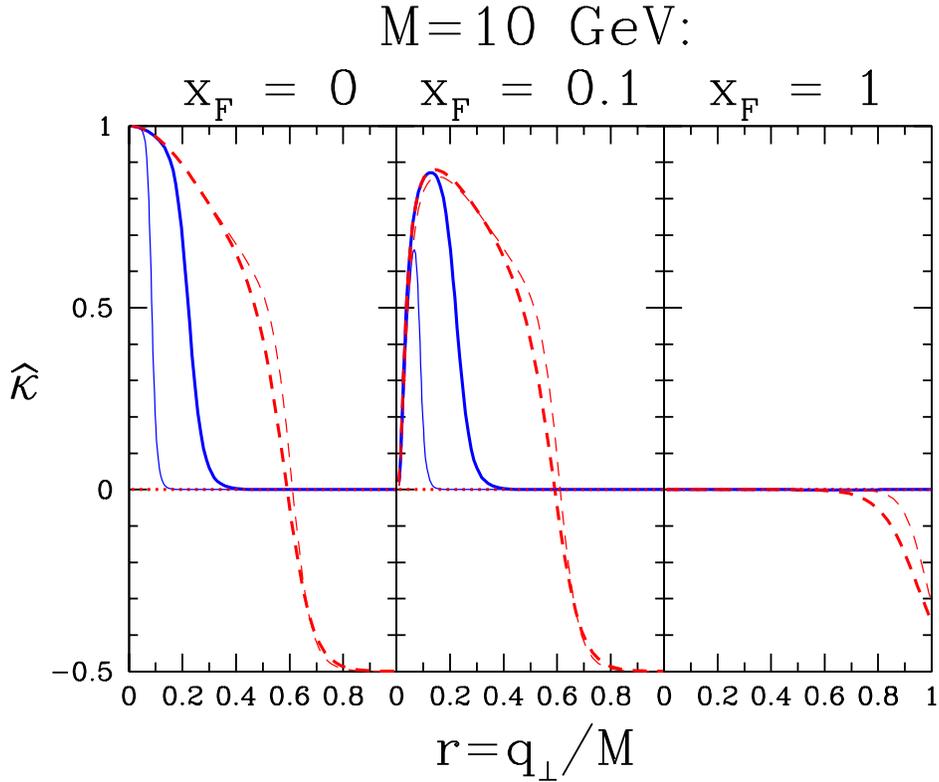,width=14.cm}
\vspace{-2.5cm}
\caption[dum]{\label{hatkappa_comp}
        Illustration of the dependence of our prediction of the Lam-Tung parameter 
        $\hat\kappa = \frac{1}{4} (1-\hat\lambda -2\hat\nu )$ on the choice of $n_f$ 
        and $\Lambda$, for fixed $M=10$~GeV and various values of $x_F$. 
      The thicker lines correspond to our default choice $n_f=3$ and
      $\Lambda=\Lambda^{(3)}_{\rm\overline{MS}}=0.346$~GeV, whereas the 
      thinner ones correspond to $n_f=1$ and
      $\Lambda=\Lambda^{(1)}_{\rm\overline{MS}}=0.241$~GeV (other notations as in 
      Figs.~\ref{coeff_funct_part_7} and \ref{coeff_funct_part_10}).}
\end{center}
\end{figure}

Strictly speaking, we should take $n_f=1$ and a corresponding $\Lambda
=\Lambda^{(1)}_{\rm\overline{MS}}$ value for the calculation of the
effective coupling parameters $b$, Eq.~(\ref{beff}), and $\xi^{\rm
  (I)}$, Eq.~(\ref{xieff_i}), of the instanton contribution to the
helicity structure functions, since our instanton-induced process
corresponds to (the unrealistic case of) one massless flavour, the
other flavours being integrated out.  In the numerical results shown
in Figs.~\ref{coeff_funct_part_7} and \ref{coeff_funct_part_10} we
have chosen, instead, $n_f=3$, and
$\Lambda=\Lambda^{(3)}_{\rm\overline{MS}}=0.346$~GeV, for the
calculation of the effective coupling parameters $\xi^{\rm (pt)}$,
$\xi^{\rm (I)}$, and $b$.  This value of $\Lambda$ corresponds --
according to the standard three-loop perturbative flavour reduction --
to an $n_f=5$ value $\Lambda^{(5)}_{\rm\overline{MS}}=0.219$~GeV,
leading to a running QCD coupling $\alpha_s(m_Z)=0.119$ at the
$Z$-boson mass~\cite{Ringwald:1999jb}.  As illustrated in
Fig.~\ref{hatkappa_comp}, our results for the coefficient functions
are not largely affected if we choose instead the nominal value
$n_f=1$ and a corresponding value for the $\Lambda$ parameter,
$\Lambda^{(1)}_{\rm\overline{MS}}= 0.241$~GeV.  This value was
obtained by a linear interpolation between the central values found in
recent lattice investigations for $n_f=0$,
$\Lambda^{(0)}_{\rm\overline{MS}}=0.237$~GeV~\cite{Capitani:1998mq},
and $n_f=2$,
$\Lambda^{(2)}_{\rm\overline{MS}}=0.245$~GeV~\cite{DellaMorte:2004bc}.
Again, such details cancel to a great extend in the ratios of
structure functions.  This also refers to the dependence on the 
renormalization scale $\mu_r$, for
which we have chosen $M$ in the numerical results presented in
Figs.~\ref{coeff_funct_part_7} and \ref{coeff_funct_part_10}.

\subsection{\label{hadron} Hadron level}

Since we have to deal with collisions of hadrons, the partonic
Mandelstam variables $s,t,u$ in Eqs. (\ref{wsigma})-(\ref{w4}) are not
observable. Firstly, we have to calculate the tensor (\ref{wmunu}) on
the hadron level which involves a folding with the usual parton
distributions, see e.g.  Ref.~\cite{Lam:1978zr},
\begin{eqnarray}
 W_{\mu\nu}(S,T,U)&=&
\frac{16\pi}{3}
\int
\frac{{\rm d}x_1}{x_1}\frac{{\rm d}x_2}{x_2}\,
S \,
\sum_{i}w^{\,q_i}_{\mu\nu}(s,t,u)\,
\lf(q_i(x_1) \bbar{q}_i(x_2)+
\bbar{q}_i(x_1) q_i(x_2)\rg) 
\label{wiha}
\\
&& \hspace*{-2.3cm}=
\frac{8\pi}{3}
\int
\frac{{\rm d}x_1}{x_1}\frac{{\rm d}x_2}{x_2}\,
\delta\lf(\frac{s+t+u-M^2}{S}\rg) 
\sum_{i}w^{\prime\,q_i}_{\mu\nu}(s,t,u)\,
\lf(q_i(x_1) \bbar{q}_i(x_2)+
\bbar{q}_i(x_1) q_i(x_2)\rg),
\label{wiha1}
\end{eqnarray}
where the flavour dependence of $w_{\mu\nu}$ is given by the relative
charge $e_{q_i}$ in $\xi^{\rm (I)}$ (\ref{xieff_i}) and $\xi^{\rm
  (pt)}$ (\ref{xieff_pt}). Note that the second equation (\ref{wiha1})
only holds for one parton in the final state, whereas the first equation is
applicable for the general partonic tensor (\ref{wmunu}).  The factors
entering the hadronic tensor (\ref{wiha1}) are fixed in such a way
that the tensor fits with the one defined in
\cite{Lam:1978zr}\footnote{Constant factors are for our purposes
  actually not important since we are only interested in ratios of
  functions $W_i$.}. We have to project the hadronic tensor
(\ref{wiha1}) now on the hadron momenta $K_1$ and $K_2$ to get the
accessible hadronic structure functions $W_i$,
\begin{equation}
 W_{\mu\nu}=-\tilde{g}_{\mu\nu}\*W_1+\tilde{K}_{\mu}\*\tilde{K}_{\nu}\*W_2
 -{\tilde{K}_{\mu}\tilde{k}_{\nu}
 +\tilde{K}_{\nu}\tilde{k}_{\mu}\over 2}\*W_3
 +\tilde{k}_{\mu}\*\tilde{k}_{\nu}\*W_4\,.
\label{wmunu_p}
\end{equation}
Here we have defined, similar to the partonic case discussed before,
the vectors, $K=K_1+K_2$, $k=K_1-K_2$,
$\tilde{K}_\mu=\tilde{g}_{\mu\nu}\,K^{\nu}/\sqrt{S}$ and
$\tilde{k}_\mu=\tilde{g}_{\mu\nu}\,k^{\nu}/\sqrt{S}$.  Note the
differences between the partonic momenta $k^\prime$, $K^\prime$ and
the hadronic ones $k$, $K$ in the hadronic tensor (\ref{wmunu_p}).
Due to the different projections on the hadron level the hadron
structure function $W_i$ is a linear combination of foldings of the
four partonic functions $w^\prime_i$ with the parton distributions.
Using the partonic functions (\ref{w1}) - (\ref{w4}) for the
instanton-induced contribution and Eqs.~(\ref{w1_pt}) - (\ref{w4_pt})
for the perturbative one, we have now everything at hand to calculate
the observable angular distributions (\ref{lambda}) - (\ref{nu}) on
the hadron level.

For fixed $M$, the angular distributions on the partonic level depend only
on $r$ and $x_F$, but they are independent of the center-of-mass (c.m.) energy
$\sqrt{s}$, because the latter is fixed due to the relation $s+t+u=M^2$. 
On the hadron level, however, the momentum fractions $x_1$ and $x_2$ are variable, 
and the angular distributions depend, correspondingly, for fixed $M$,  on the hadronic c.m.  
energy $\sqrt{S}$. 
Furthermore, the variable $x_F$ has to be replaced by the
hadronic one,
\begin{equation}
 X_F=\frac{T-U}{S}\,,
\end{equation}
which can be interpreted as the longitudinal photon-momentum fraction
with respect to the momentum of the hadron $h_1$.

\begin{figure}
\centering
\includegraphics*[width=17.cm]{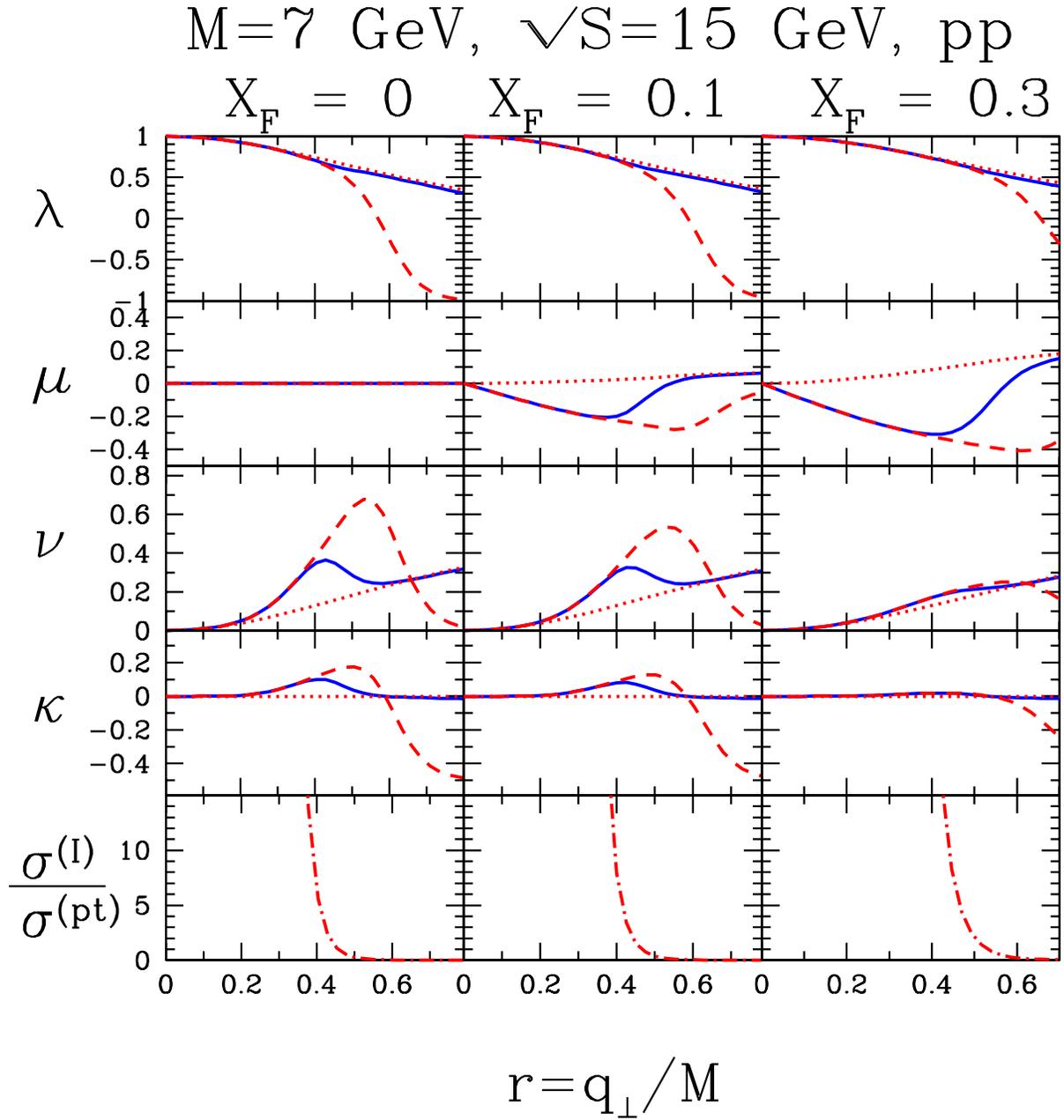}

\caption{\small\label{lammunu}
The plot shows the angular structure
functions similar to Fig.~\ref{coeff_funct_part_7} but on hadron level.
Therefore an integration over parton distributions, e.g. for the proton, is included. 
Due to the variable momentum fractions $x_1$ and $x_2$ one has to
specify an additional kinetic variable, e.g. the  c.m. energy
$\sqrt{S}$.}
\end{figure}

Figure~\ref{lammunu} shows the resulting angular distributions for
proton-proton collisions at $\sqrt{S}=15\;{\rm GeV}$ for $M=7\;{\rm
  GeV}$, $X_F=0,0.1,0.3$, and varying values of $r=q_\perp/M$.  For
the renormalization scale we have chosen $\mu_r=M$ and for the parton
distributions the CTEQ6 dataset
\cite{Pumplin:2002vw}\footnote{Actually a consistent treatment of
  instanton induced effects requires also parton distributions
  including instanton-induced parton evolution. Since this
  modification enters the perturbative and instanton contribution in
  the same way this effect would change the angular distribution only
  in sub-leading order of instanton perturbation theory.}.

\begin{figure}
\centering
\includegraphics*[width=17.cm]{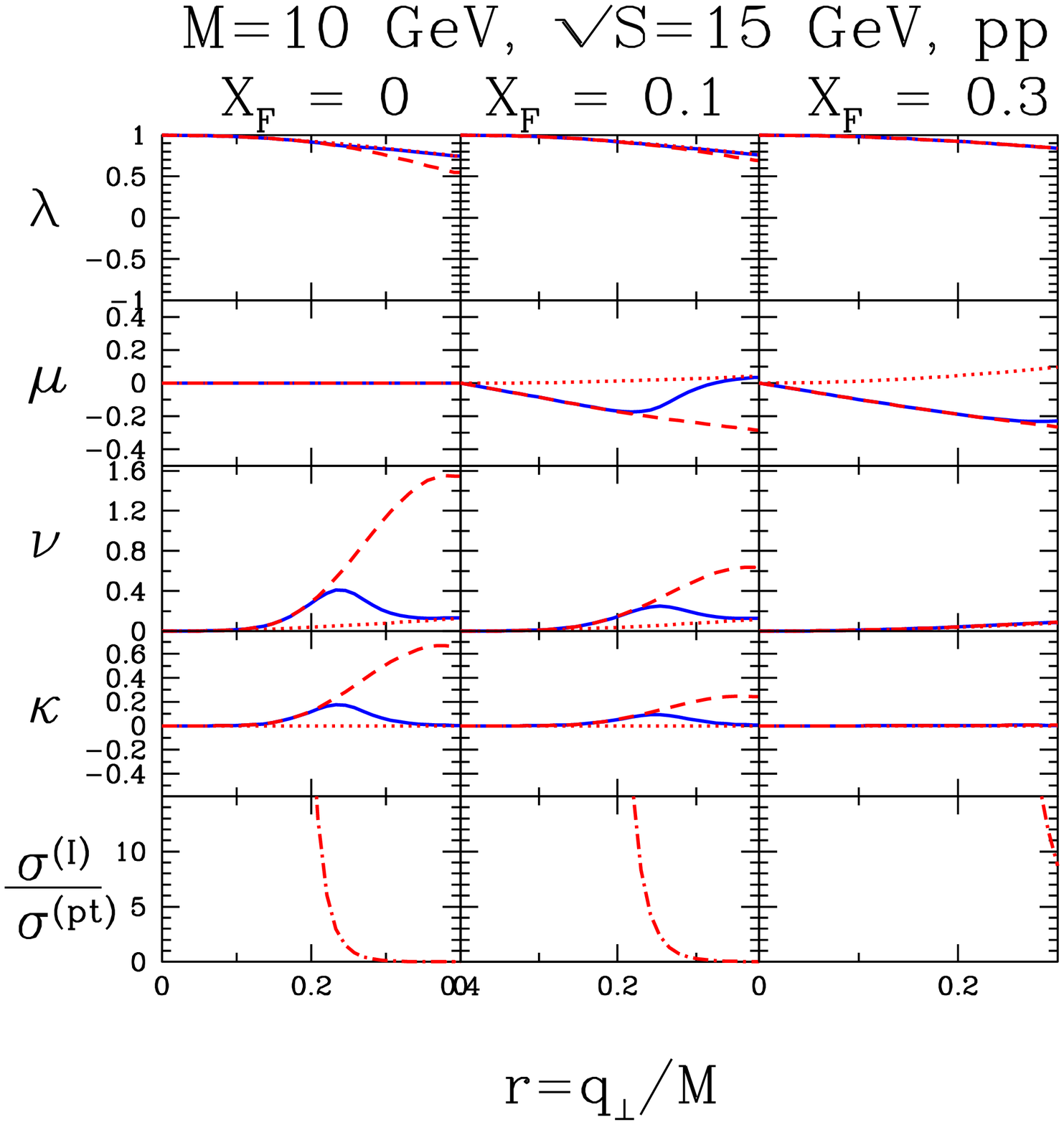}

\caption{\small\label{lammunu_10}
      Same as in Fig.~\ref{lammunu}, but for $M=10$~GeV.
}
\end{figure}

The main difference to the partonic quantities, shown in 
Fig.~\ref{coeff_funct_part_7}, concerns the $x_F$ resp. $X_F$
dependence. The strong $x_F$ dependence is smeared out on the hadron
level. This smearing leads in particular to a suppression of the
instanton-induced effect at small $X_F$. This has nothing to do with
the parton distribution functions entering the hadronic angular
distribution. Actually, since the angular distributions are {\it
  ratios} of two foldings with parton distribution functions, their
dependence on the type of hadrons in the initial states is rather
weak.  The difference is just a consequence of the $x_{1,2}$
dependence of the partonic Feynman variable $x_F$ that leads to the
smeared out $X_F$ behavior after integrating over $x_{1,2}$. For
similar reasons, the $M^2$ dependence of the angular distribution in
the instanton background is stronger than on the parton level, since
for smaller ratios $M^2/s$ also smaller values of $x_F$ contribute,
see Fig.~\ref{lammunu} and \ref{lammunu_10}; but note that due to
kinematical reasons only smaller values of $r=q_\perp/M$ are
accessible for larger ratios $M^2/s$.

The fiducial region of instanton perturbation theory on the parton
level $\sqrt{-t},\sqrt{-u}, M>{\mathcal Q}_{\rm min}\approx (4-6)\;{\rm
  GeV}$ can be mapped on the hadronic variables.  One can check that
theses relations are fulfilled for all $x_1$ and $x_2$ for large
enough values of $M$ and $q_\perp$, namely $M\ge{\mathcal Q}_{\rm
  min}$ and $q_\perp\gwig
  {\mathcal Q}_{\rm min}$. Also on the hadron level our results should hold
for even smaller $q_\perp$, since the uncertainty towards smaller $q_\perp$
drops out in the ratios of the angular distributions, see the
discussion in Sect.~\ref{partonlevel}. 


\subsection{\label{gluons} More partons in the final state} 
As already mentioned in the introduction, the discussed
instanton-induced sub-process $q\bar{q}\stackrel{I}{\to}\gamma^\ast g$
with only one gluon and no quarks in the final state is quite
instructive since it contains already the basic non-trivial feature of
the instanton-induced Drell-Yan process, namely the helicity flip of
the quarks in the initial state, which is related to the 
chirality violation and is essentially responsible for the violation
of the Lam-Tung relation \cite{Brandenburg:1993cj,Boer:2004mv}. But the rate of this
asymmetry induced by instantons was certainly underestimated in the
previous sections since it is well known that the resummation of the
events with an arbitrary number of final-state gluons leads to a large
enhancement which eats up at least partially the suppression of the
instanton-induced process $q\bar{q}\to \gamma^\ast g$. 
In addition, also the number of involved quarks in
the subprocess is not realistic, see Sect.~\ref{dis2drellyan}.

A complete calculation of the angular distribution for this general
instanton-induced process is beyond the scope of this paper and will be
attempted in the future. Let us roughly sketch the general features of the
complete process. Whereas in perturbative processes additional
final-state gluons are certainly suppressed by an order of $\alpha_s$, 
every additional gluon in an instanton background leads to an {\it
  enhancement} of the order $1/\alpha_s$.  Summing over all processes
with an arbitrary number of gluons $n_g$ leads to an exponentiation of
the inverse coupling constant. The resulting factor, combined with the
tunneling factor, $\exp[-4\pi/\alpha_s]$ (cf. e.g. Eq.~(\ref{xieff_i})), 
can be written as $\exp[-4\pi/\alpha_s \,F(x^\prime)]$. 
Here, the Bjorken scaling variable $x^\prime = \mathcal{Q}^2/(\mathcal{Q}^2+M_X^2)$ appears, 
where $\mathcal{Q}$ is the relevant momentum transfer and $M_X$ is the invariant
mass of the produced partonic final state. 
The so-called holy-grail function $F(x^\prime)$ \cite{McLerran:1989ab} is normalized to one for $x^\prime=1$ and decreases 
towards smaller $x^\prime$ and therefore larger $M_X$. Let us mention that in the
electroweak theory, where the coupling constant is much smaller, this
mechanism is absolutely necessary for the process eventually becoming observable
in the high energy limit~\cite{Aoyama:1986ej}.

For the process discussed in the present paper, $\mathcal{Q}^2$ is given
by the partonic quantities $-t,-u$ or $M^2$, whereas, in general,
$M_X^2=(k_1+k_2-q)^2$.  Therefore, the integrands of the functions
$W_i$ (\ref{wiha1}) involve a factor
$\exp[-4\pi/\alpha_s\,F(x^\prime)]$.  For $M_X^2=0$ ($x^\prime=1$),
and therefore also for the process with one final-state gluon, the
factor $\exp[-4\pi/\alpha_s]$ in Eq.~(\ref{xieff_i}) is recovered.
For positive $X_F$, the smallest $x^\prime$ is given for
$\mathcal{Q}^2=-t$,
\begin{equation}
 x^\prime=\frac{-t}{-t+M_X^2}=\frac{-t}{s+u-M^2}=\frac{-2 M^2/S +  x_1(
   \sqrt{4M^2/S(1 + r^2) + X_F^2}-X_F)}
{2 x_2 x_1 -x_2(\sqrt{4M^2/S(1 +r^2)+X_F^2}+X_F)}\,.
\end{equation}
It is easy to check that $x^\prime$ rises slightly with $r$
and decreases towards the largest accessible values of $X_F$.
Therefore, we can conclude that the instanton-induced effect in
$\nu$ and $\kappa$ (see Figs.~\ref{coeff_funct_part_7},
\ref{coeff_funct_part_10}, \ref{lammunu} and
\ref{lammunu_10}) will shift to slightly smaller $r$.
Furthermore, we expect a significant enhancement of the instanton
effect for larger $X_F$. Correspondingly, the suppression of the simplest
instanton-induced process at large $X_F$, which we observed before, might be
compensated. Note that the applicability of instanton
perturbation-theory now requires in addition a cut $x^\prime\gwig x_{\rm
  cut}$, where $x_{\rm cut}$ is approximately 0.35, see Ref.~\cite{Ringwald:1999ze}. 
One can check that this requirement can be
fulfilled for all $x_1$, $x_2$  and $r$ as long as $X_F$ is not too 
large, or for all $X_F$ for large enough ratios  $M^2/S$.

Beside the discussed instanton-induced multi-gluon process also other
perturbative processes may contribute. Firstly, we have not taken into
account an enhancement of perturbative contributions due to soft gluon
resummation \cite{Chiappetta:1986yg,Boer:2006eq} at small $q_\perp$
since the instanton-induced contribution that we have calculated is
not reliable in this region anyway. In higher order $\alpha_s$ also
new processes contribute to the angular distribution which lead to a
small violation of the Lam-Tung relation already in the purely
perturbative framework \cite{Mirkes:1994dp}. For small tranverse
momenta $q_\perp$ the usual factorization is not reliable anymore and
transverse parton momentum distributions become important. However, as
already mentioned in the introduction, higher order contributions,
soft gluon effects and parton transverse momentum are not able
to explain the observed strong violation of the Lam-Tung
relation \cite{Mirkes:1994dp,Chiappetta:1986yg,Boer:2006eq}.

\section{\label{conclusions}Conclusions}
We have calculated the angular distribution of the produced leptons in
hadron hadron collision in an instanton background. It turns out that,
for large enough photon virtualities $M^2$ and transverse photon
momenta squared $q_\perp^2$, only small instantons contribute. Therefore, the
instanton-induced contribution is fiducially calculable in this
kinematic region using techniques of instanton perturbation theory.
The most remarkable property of the resulting angular distribution is
the violation of the Lam-Tung relation which is conserved to very high
accuracy in usual perturbation theory, but violated in experiments.
This effect is a direct consequence of chirality violation in the
background of QCD instantons which leads to a non-trivial spin-density
of the quark-antiquark pair in the initial state as it has been argued
in \cite{Boer:2004mv}.  Therefore, lepton pair production in hadron
collisions is potentially a very good testing ground for
instanton-induced processes: the 
violation of the Lam-Tung relation is 
reliably calculable in instanton perturbation theory and absent in usual perturbation theory.

We restricted ourselves to the simplest partonic subprocess $q\bar{q}\stackrel{I}{\to}\gamma^\ast+g$. 
Since the inclusion of the more realistic general processes
$q\bar{q}\stackrel{I}{\to}\gamma^\ast+(n_f-1)q\bar{q}+n_g g$ was beyond the scope
of this paper, we cannot compare our results directly with the
available data. 
However, the small violation of the Lam-Tung relation on the 
hadron level arising from the simplest partonic process is already quite promising, 
notably in view of
the expectation that additional gluons lead to a substantial enhancement of the
instanton-induced effect, as known from analyses of the related processes 
in deep-inelastic scattering and from the general arguments presented in this paper. 

Finally, let us mention that, beside further theoretical efforts, more
experimental data are required for testing instantons in the angular
distribution of produced leptons at a hadron collider.  Fortunately,
there are new medium energy projects under way 
that are also dedicated to study the Drell-Yan process, 
e.g. at the forthcoming facilities GSI-FAIR~\cite{PAX} 
and J-PARC~\cite{JPARC}.\footnote{At J-PARC a proton beam of 50~GeV
  ($\sqrt{S}\approx 10\;{\rm GeV}$) will be used for fixed target
  experiments and at the FAIR experiment a 29~GeV antiproton beam will be
  available for fixed target experiments or collisions with low energy
  protons ($\sqrt{S}\approx 6 - 15\;{\rm GeV}$). Clearly, 
  proton-antiproton collisions are perfectly suited for studying
  Drell-Yan since the rate is higher as in proton-proton collisions.}
Experiments at RHIC may also give further information on lepton pair production. 
In general, it seems that fixed target experiments are especially well suited
for our purposes, since on the one hand the involved momenta are
smaller and on the other hand the luminosities are larger. Therefore, a
huge amount of lepton pairs should be observable which is absolutely
necessary for reconstructing a whole angular distribution.  

\section*{Acknowledgments}

One of us (A.R.) would like to thank Sven Moch and Fridger Schrempp
for sharing their insights into QCD instanton-induced hard scattering
processes with him.  A.U. would like to thank Fridger Schrempp for
countless illuminating discussions about instantons and beyond.  A.B.
and A.U. would like to thank also Otto Nachtmann and Dani\"el Boer for
numerous discussions about lepton pair production in hadron collisions
as a testing ground for perturbative QCD.  We thank also Markus Diehl
for valuable information on future experiments on Drell-Yan
production.  The work of A.B. was partially supported by a Heisenberg
grant of the Deutsche Forschungsgemeinschaft.  The work of A.U. was
supported by the research program of the ``Stichting voor
Fundamenteel Onderzoek der Materie (FOM)'', which is financially
supported by the ``Nederlandse Organisatie voor Wetenschappelijk
Onderzoek (NWO)''.

\end{document}